\documentclass[9pt]{article}
\usepackage[utf8]{inputenc}
\usepackage{amsmath,amsfonts,amssymb,latexsym}
\usepackage[T1]{fontenc}
\newtheorem{satz}{Theorem}[section]

\newtheorem{defi}[satz]{Definition}

\newtheorem{bem}[satz]{Remark}
\newtheorem{lemma}[satz]{Lemma}
\newtheorem{koro}[satz]{Corollary}
\newtheorem{conclusion}[satz]{Conclusion}
\newtheorem{ob}[satz]{Observation}

\newcommand{\mcal}{\mathcal}
\newcommand{\mbf}{\mathbf}

\newcommand{\tit}{\textit}

\newcommand{\beq}{\begin{equation}}
\newcommand{\eeq}{\end{equation}}

\begin{document}
\thispagestyle{empty}
\begin{center}
\vspace*{1.0cm}
{\Large{\bf The Rigorous Relation between Rindler\\ and Minkowski Quantum Field Theory\\ in the Unruh Scenario }}
\vskip 1.5cm
{\large{\bf Manfred Requardt}}

\vskip 0.5cm

Institut fuer Theoretische Physik\\
Universitaet Goettingen\\
Friedrich-Hund-Platz 1\\
37077 Goettingen \quad Germany\\
(E-mail: requardt@theorie.physik.uni-goettingen.de) 

\end{center}
\begin{abstract} Traditionally the physics of the Unruh effect, i.e. the q.f.t. in the wedges $W_R$ or $W_L$ in Minkowski space is related to the physics in the Rindler Fock space, which is a proplematical strategy. In a careful analysis we show that the correct dual q.f.t. lives rather in the thermal Rindler Hilbert space and turns out to be unitarily equivalent to the corresponding Minkowski space theory in contrast to the Rindler Fock space theory. We show in particular that in thermal Rindler Hilbert space a new sort of objects occurs, viz., quasi-particle/hole creation/annihilation operators of thermal Rindler quasi-particles and holes, which do not have a pendant in Rindler Fock space. The ordinary Rindler particle operators are certain temperature dependent superpositions of these more fundamental operators. These new objects play a crucial role in this duality and via the unitary equivalence do have their counterparts in Minkowski q.f.t. 
\end{abstract}
\newpage 
\section{Introduction}
While there exist a host of papers, even reviews, dealing with the so-called \tit{Unruh effect}, there are in our view and in the view of some other colleagues, a number of open questions which are worthwhile to be addressed and carefully discussed. This holds the more so as some of them are apparently not even regarded as problems at all. We think that the Unruh effect is one of the not so frequent cases in physics where technical aspects really do matter.  

Most of the papers and reviews ( as there exist such a large number of papers and as we do not intend to write another review we cite only a few , as we think, representative references, see e.g. \cite{C},\cite{T},\cite{W1},\cite{S}) are based to a large degree on the methods of the fundamental paper by Fulling (\cite{F}) about the non-equivalence of field quantization in Minkowski respectively Rindler space-time and the framework developed a little bit later by Unruh,\cite{U}). While the emphasis of Unruh's approach lies more on the Minkowski regime (at least according to his own remarks (see \cite{FU}), in practically all these papers a central piece is the calculation of the Bogoliubov coefficients relating the two different quantization schemes. 

One should note that, while the respective Fock space representations of the observable algebras  are not even unitarily equivalent (as we will show in the following), the creation/annihilation operators of the KG-field in the Minkowski and Rindler representation are frequently assumed to be related by number-valued Bogoliubov coefficients instead of operators. This is an ambiguous point we will address in detail. This problematic assumption presumably derives from the following idea. We are given the K.G.-quantum field $\phi(x)$. Classically we can of course choose another parametrization , e.g. Rindler coordinates and freely switch between these two coordinatizations. However, in the quantized case the quantum field is defined on a particular Hilbert space and $\phi(x)$ defined on Minkowski Hilbert space cannot automatically be identified with $\phi(\rho,\eta)$ defined on Rindler Fock space. The same applies to the different mode expansions.

One then usually proceeds by (formally) expressing the Minkowski vacuum vector as a thermal density  matrix in the Hilbert space of the right Rindler wedge and the Minkowski space annihilation operators of the KG field as a complicated superposition of Rindler creation/annihilation operators or vice versa. While it is sometimes remarked in passing that this is perhaps not completely correct it is nevertheless assumed for convenience. This is another point we will analyse.
\begin{bem} We discussed this point briefly in e.g. \cite{Requ1} observation 3.2. As the generator of time evolution in Rindler space is the generator of Lorentz boosts, the corresponding Hamiltonian has a continuous spectrum while the Hamiltonian in a representation where the vacuum can be represented as a density matrix is the logarithm of the density operator and has therefore a discrete spectrum.
\end{bem}
 To sum up, we think the main problem of a large body of work on the Unruh effect is that there is a permanent mixing and, sometimes, even identification of two views, the Minkowski and the Rindler viewpoint and, correspondingly, a mixing of states as expectation functionals and Hilbert space vectors in the (possibly) inequivalent representations of the field or observable algebra. We will show that this is a point that matters because it is not at all clear to what extent it is allowed to relate the various expressions, belonging to different and possibly inequivalent representations of the K.G. field algebra to each other.

There exists a completely different relatively abstract and more mathematical approach based on the work of Bisognano-Wichmann, the original intention of which was not to discuss the Unruh-effect (\cite{Wich1},\cite{Wich2}). While the results of Bisognano-Wichmann refer to more general phenomena in axiomatic quantum field theory they are mathematically much more involved and less explicit and transparent in the concrete Unruh-situation. On the other hand, the original treatment employs standard methods of quantum field theory and provides quite explicit formulas.

A more mathematical discussion of the complex of questions related to the so-called \tit{KMS-property} of the states under discussion in this context is for example given in \cite{Kay1} and \cite{Kay2}. A little bit closer to the proper Unruh effect, but still starting from the Bisognano-Wichmann framework is the paper by Sewell (\cite{Sewell}).

One can perhaps resume this short summary by stating that there exists, on the one hand, a rigorous but relatively abstract analysis (the approach inspired by the work of Bisognano-Wichmann) with the physical implications being less explicit and, on the other hand, a direct approach in the spirit of ordinary quantum field theory (starting from the work of Fulling and Unruh) providing a lot of physical details and insight but using methods which are sometimes perhaps a little bit problematical.

In the following we want to develop a third approach which combines the merits of both approaches mentioned above. In this process we undertake to discuss and clarify various points which are, in our view, perhaps a little bit sloppily treated in the standard treatment. Furthermore we try to add a number of, as we think, interesting details which we have not found in the literature known to us.

More specifically, we want to address the following points:\\
i) The observation that the Minkowski vacuum cannot be represented as a density matrix within the Rindler Hilbert space framework is actually connected with some deeper and interesting properties of  v.Neumann algebras of observables in both scenarios (that is, Minkowski or Rindler in the, say, right wedge case $W_R$).\\
ii) We show that one can almost completely avoid the reference to the Rindler Fock space based on the Rindler vacuum vector (the latter is presumably not easy to realize physically anyhow) and can rather perform a large amount of the necessary  calculations within the thermal Rindler Hilbert space with the observable algebras of the left/right wedge turning out to be unitarily equivalent to the corresponding algebras in Minkowski Hilbert space. A fortiori, we show that the thermal (KMS) property of the Minkowski vacuum can be expressed with the help of the two-point function and expressing the Minkowski coordinates as functions of the corresponding Rindler coordinates. These Rindler coordinates are then analytically continued to show that the Minkowski vacuum vector fulfills the so-called KMS-condition.\\
iii) We show that one should replace in a correct treatment the ordinary Rindler Fock-space creation/annihilation operators by a superposition of thermal quasi-particle/hole creation/annihilation operators acting on the (infinite volume) thermal state (the thermal Rindler vacuum) which corresponds to the Minkowski vacuum in the Rindler framework. We show to what extent this correspondence holds in a rigorous sense. This will be discussed in detail in sections 4 and 5.  Furthermore, we show explicitly how all these (thermal) objects can be mapped into corresponding objects in Minkowski Hilbert space via the unitary equivalence we will prove in the following.    
\section{Observable Algebras on $W_R$ as a Subset of Minkowski or Rindler Space}
In this section we want to compare the algebras of observables generated by a free hermitean scalar field $\phi (t,x)$ on $W_R$, i.e., the (open) right wedge given by
\beq (t,x)\quad \text{with}\quad |t|< x             \eeq
For convenience (as we are dealing mostly with questions belonging to quantum field theory) we choose the Minkowski metric as
\beq ds^2=dt^2-dx^2 \quad \text{with}\quad c=1    \eeq
and we restrict ourselves to the case of two dimensions. While the extension to e.g. four dimensions needs some extra calculations it is nevertheless straightforward and our choice is entirely sufficient concerning matters of principle.

We begin with the open wedge $W_R$. Rindler coordinates $(\rho,\eta)$ are given by
\beq  t=\rho\cdot \sinh \eta  \quad x=\rho\cdot \cosh \eta\quad 0<\rho<\infty\;,\;-\infty<\eta<+\infty    \eeq
Sometimes the further coordinate transformation
\beq t=a^{-1}e^{a\xi}\sinh a\tau\quad x=a^{-1}e^{a\xi}\cosh a\tau    \eeq
is made with
\beq \rho=a^{-1}e^{a\xi}\quad ,\quad \eta=a\tau   \eeq
where $a$ is a positive constant. In this coordinate system the world line with $\xi=0$ has a (proper) acceleration $a$ and $\tau$ is the proper time of an observer on this world line.

In these coordinates the metric reads
\beq ds^2=\rho^2d\eta^2-d\rho^2   \eeq
or
\beq ds^2=e^{2a\xi}(d\tau^2-d\xi^2)   \eeq
One sees that the latter coordinate system has certain advantages concerning quantization as it is conformal to the Minkowski metric, but we usually prefer the former system.

At the moment we are not so much interested in constructing creation and annihilation operators (as it is done in e.g. \cite{F} and \cite{U}) but we rather want to study the field algebras or, more specifically, the algebras of observables (i.e. the bounded operators constructed from the fields $\phi(t,x)$ or $\phi(\eta,\rho)$, in other words, polynomials in the smeared fields
\beq \phi(f):=\int \phi(t,x)f(t,x)dtdx\quad\text{or}\quad \phi(g):=\int \phi(\eta,\rho)g(\eta,\rho)d\eta d\rho   \eeq
with smooth test functions $f,g$ having compact support in the open $W_R$ and certain limit operators or employing the respective spectral measures to get bounded operators.

We see that as long as we remain in the interior of $W_R$ the correspondence between $(t,x)$ and $(\eta,\rho)$ is bijective, i.e., to each test function $f(t,x)$ belongs a test function $f'(\eta,\rho)$ and vice versa. This entails that from such smeared field operators we can construct certain
 (open) algebras of observables which can be mapped bijectively and linearly onto each other by an \tit{algebraic isomorphism}. In order to be able to draw some strong conclusions we have to introduce an appropriate operator topology.

We now assume that the one algebra, $\mcal{A}_R$, which uses Minkowski coordinates, is a subalgebra of the field or observable algebra $\mcal{A}_M$ of the Minkowski Hilbert space, while the other one, $\mcal{B}_R$, is built from the field operators belonging to the Rindler Fock space. Up to now all elements of the algebras are localized in the interior of $W_R$. We want to make both algebras into so-called v.Neumann algebras by closing them in, for example , the weak operator topology, that is, we want to include limit elements $A$  so that there exists a sequence $A_n\in \mcal{A}_R$ or $\mcal{B}_R$ with
\beq \lim_{n\to\infty} (\psi|A_n\phi)=(\psi|A\phi)\quad\text{for all pairs}\quad (\psi,\phi)   \eeq
taken from the respective Hilbert spaces.
\begin{bem} We note in passing that v.Neumann algebras happen to be closed in other operator topologies. Another important topology in this context is for example the $\sigma$-weak topology. That is, $A_n$ converges to $A$ in the $\sigma$-weak topology if it holds that for all pairs of sequences $(\psi_i,\phi_i)$ with 
\beq \sum_i |\psi_i|^2<\infty\quad \sum_i |\phi_i|^2<\infty   \eeq
it holds 
\beq \lim_{n\to\infty}\sum_i |(\psi_i|A_n\phi_i)| =\sum_i |(\psi_i|A\phi_i)|   \eeq
\end{bem}

We want to study what happens if we approach the boundary of the wedge $W_R$, i.e. the \tit{null planes} $t=x,x>0$ or $t=-x,x>0$. With $t-x\to 0$ we have $t^2-x^2\to 0$ hence $\rho\to 0$. On the other hand, 
\beq t/x=\tanh\eta\to 1\quad\text{implies}\quad \eta\to\infty   \eeq

In Minkowski coordinates it is not difficult to construct limit elements which are concentrated on or near the boundary, that is, for example on the null plane $t=x$. Assuming that the test functions of the sequence are concentrated in $W_R$ uniformly between the two planes 
\beq t+x=a>0\quad ,\quad t+x=b>a  \eeq
and are shrinked with $n\to\infty$ in the transversal direction with the help of test functions 
\beq h_n(t-x)\quad ,\quad \int h_n(s)ds=1\quad ,\quad\text{for example}\quad h_n(s)=n\cdot h(ns)   \eeq 
we can construct limit elements which are concentrated on or near the boundary, that is, which live on the null plane $t=x$. We omit the technical details which can be found in e.g. \cite{Sewell},\cite{Leutwyler},\cite{Schlieder},\cite{Rohrlich} or \cite{Nakanishi}. In this way we complete the non-complete algebra $\mcal{A}_R$  and make it into a weakly closed v.Neumann algebra. We come back to this point in section 4.2.

There is evidently no problem because Minkowski coordinates happen to be completely regular along the null planes. We now try to perform a similar construction in the Rindler regime. By trying to concentrate the support of fields or (more generally) observables near the null plane  $(t=x)$, we see that the corresponding $\eta$-coordinates wander away towards $\infty$. This implies that the corresponding elements from $\mcal{B}_R$ do not converge to (non-vanishing) weak limit elements.
\begin{bem} For sufficiently localized Hilbert vectors $\psi$ or $\phi$ within $W_R$  we see that the scalar product $(\psi|A_n\phi)$ will go to zero when $A_n$ moves to $\infty$ in Rindler coordinate space (for the technical details see section 4.2).
\end{bem}
We hence arrive at the important conclusion that the weak closures of $\mcal{A}_R,\mcal{B}_R$, i.e., $\overline{\mcal{A}}^w,\overline{\mcal{B}}^w$ are different.
\begin{conclusion} The weak closures of  $\mcal{A}_R,\mcal{B}_R$, i.e., $\overline{\mcal{A}}^w,\overline{\mcal{B}}^w$, contain different sets of limit elements. That is, while  $\mcal{A}_R,\mcal{B}_R$ are algebraically isomorphic, this is no longer the case for  $\overline{\mcal{A}}^w,\overline{\mcal{B}}^w$.
\end{conclusion}

This has interesting consequences as there exist a number of deeper results if  $\overline{\mcal{A}}^w,\overline{\mcal{B}}^w$ were algebraically isomorphic as v.Neumann algebras. In the following we mention such results which can e.g. be found in \cite{Bratteli}.
\begin{satz} If $\mcal{M},\mcal{N}$ are two v.Neumann algebras and $\Phi$ an algebraic $\ast$-isomorphism from $\mcal{M}$ to $\mcal{N}$ then $\Phi$ is $\sigma$-weakly continuous in both directions.
\end{satz}
The so-called \tit{normal states} are states on a v.Neumann algebra given on some Hilbert space which are continuous in the $\sigma$-weak topology on the algebra. It is well known that they are given by vector states and density matrices. Due to the property of $\Phi$ in the preceding theorem we have immediately:
\begin{satz} With $\Phi$  $\sigma$-weakly continuous each normal state $\omega$ on $\mcal{M}$ defines uniquely a normal state $\omega\circ\Phi^{-1}$ on $\mcal{N}$ and vice versa. (See \cite{Bratteli} theorems 2.4.23,2.4.26)
\end{satz} 
We see from these results that the possibility that the Minkowski vacuum vector can be considered as a density matrix in the Rindler representation appears to be related in a subtle way to the behavior of the corresponding observable algebras under physically motivated topological closure operations near the event horizon (in the Rindler case).
\section{The Minkowski Vacuum restricted to the Rindler Wedge as a Thermal (KMS) State}
In this section we will show that the Minkowski vacuum restricted to the Rindler wedge $W_R$ is a thermal (KMS) state without using the properties of the Rindler Hilbert space. That is, according to our philosophy, we show that quite a lot can already be calculated within the scenario of the Minkowski Hilbert space. 

As we have to deal with an infinite quantum system and according to the observations made in the preceding section we avoid to discuss thermality with the help of thermal (Gibbsian) density matrices as it is typically done in most of the papers discussing the Unruh effect. We will instead employ the framework which has been developed with the aim to deal with thermal system of infinite extent with the help of the so-called KMS-property (Kubo-Martin-Schwinger), see e.g. \cite{Haag} and \cite{Bratteli2}. It has been shown that the KMS-property, which holds for finite Gibbs equilibrium systems, can be extended to infinite systems and characterizes thermal equilibrium systems uniquely.

For convenience of the reader we repeat the argument for a finite Gibbs equilibrium system. We have, using the cyclicity of the trace, for two observables $A,B$:
\begin{multline} <A(t)B>_{\beta}:= Tr\,(e^{-\beta H}A(t)B)=Tr\,(e^{-\beta H}e^{iHt}Ae^{-iHt}B)=\\Tr\,(e^{iH(t+i\beta)}Ae^{-iH(t+i\beta)}e^{-\beta H}B)=Tr\,(e^{-\beta H}BA(t+i\beta)=<BA(t+i\beta)>_{\beta}
\end{multline}
\begin{bem} As $He^{-\alpha H}$ is trace class for $\alpha>0$ the function\\ $Tr\,(e^{-\beta H}BA(t+i\eta))$ is analytic for $0<\eta<\beta$ (as $\beta-\eta>0$ in $e^{-H(\beta-\eta})$.
\end{bem}

One formulation of the KMS-condition reads as follows:
\begin{defi}[KMS] For any pair of observables $A,B$ there exists a complex function $F_{A.B}$ which is analytic on the open strip $D_{\beta}=(z=t+i\eta,0<\eta<\beta$ and continuous on the closure $\overline{D_{\beta}}$. It holds
\beq  F_{A,B}(t)=<BA(t)>_{\beta}\quad ,\quad F_{A,B}(t+i\beta)=<A(t)B>_{\beta}=:G_{A,B}(t)   \eeq
Furthermore $F_{A,B}(z)$ is bounded by 
\beq ||F_{A,B}(z)|\leq ||A||\cdot ||B||   \eeq
\end{defi}
\begin{bem} Note  that in this context there do exist various minor technical points which we do not mention in order to be brief and which are discussed in the cited literature. The condition is in particular formulated for bounded observables $A,B$ while we are using in the following the unbounded scalar field operators $\phi(x),\phi(y)$ in stead of bounded observables as they lead to more transparent expressions.

At this point there exists the possibility to start from the Bisognano-Wichman framework using the modular theory. But  as this is a more abstract and perhaps less transparent  route we prefer the more direct approach descibed below. Furthermore we suppose not all readers are aquainted to this more mathematical framework.
\end{bem}

Thermal behavior of the Minkowski vacuum was observed without using the full Rindler quantization (at least to a larger part) by studying the behavior of accelerated detectors (see e.g. \cite{S},\cite{Audretsch}, or \cite{DeWitt}). However these results were restricted to particular observers (with $\rho$ fixed and Rindler time $\eta$ running) and in the end the usual Rindler mode expansion was employed to show thermal behavior. We want in the following to complement these observations with a general and direct proof of Gibbsian thermal behavior which stays completely in Minkowski space and does not invoke a discussion of the behavior of various detector models. Our idea is the following. As we deal with the free hermitean scalar Klein-Gordon (KG) field, all expressions in this model field theory can be derived from the two-point function $<\Omega|\phi(x)\phi(y)\Omega>$, that is, all higher Wigthman functions are products of such two-point functions. Hence it is sufficient to study the structure of $<\Omega|\phi(x)\phi(y)\Omega>$.

We undertake to construct a thermal state over the wedge algebra as an \tit{expectation functional} $\omega_{\beta}$, i.e. in the form $\omega_{\beta}(A)$ with $A$ some expression in the algebra of observables, by starting from the Minkowski Hilbert space expression $<\Omega|\phi(x)\phi(y)\Omega>$ with the coordinates $x,y$ restricted to $W_R$. We note that we could use a parametrisation by  Rindler coordinates instead of $x,y$ but it turns out that this is not necessary. In contrast to the ordinary Minkowski space framework we use a new time evolution together with its infinitesimal generator leading also to a new Hamiltonian.  

As is well-known, instead of the ordinary Minkowski time evolution we have to employ the representation of the Lorentz boosts in $x^1$-direction (note that we are concentrating on the two-dimensional case with $x=(x^0,x^1)$), that is, $U(\Lambda(s))$ and their infinitesimal generator $K$. These Lorentz boosts act as
\beq \Lambda(s):(x^0,x^1)\mapsto (x^0\cosh s+x^1\sinh s,x^1\cosh s+x^0\sinh s)=:x_s   \eeq
with 
\beq d/ds\Lambda(s)|_{s=0}(x^0,x^1)=(x^1,x^0)  \eeq
yielding the \tit{Killing vector field}
\beq B(x^0,x^1)=x^1\partial_{x^0}+x^0\partial_{x^1}  \eeq
From $(x^1)^2-(x^0)^2>0$ in $W_R$ we see that $B(x^0,x^1)$ is time-like.

As expected, $\Lambda(s)$ acts additively in the parameter $s$,
\beq  \Lambda(s)\Lambda(s')=\Lambda(s+s')   \eeq
This follows e.g. by using the summation formulas for $\cosh,\sinh$. In Rindler coordinates we have alternatively
\beq \Lambda(s)(\rho,\eta)=(\rho,\eta+s)   \eeq
In Minkowski Hilbert space the Lorentz boosts are algebraically represented by
\beq  U(\Lambda(s))\phi(x)U^{-1}(\Lambda(s))=\phi(\Lambda(s)x)    \eeq

In two space-time dimensions the two-point functions for the KG-field has the form
\beq  W(x-y)=<\phi(x)\phi(y)>_0=\int dk^1/(2\pi 2\omega_k)\,e^{-ik^0(x^0-y^0)+ik^1(x^1-y^1)}   \eeq
with $k^0=\omega_k=\sqrt{( k^1)^2+m^2}$. In the follwing we abbreviate $ U(\Lambda(s))$ by $U(s)$. We then have to study the expressions
\beq  <U(s)\phi(x)U(-s)\phi(y)>_0=<\phi(x_s)\phi(y)>_0=:G_ {\phi(x)\phi(y)}(s) \eeq
and
\beq <\phi(y)U(s)\phi(x)U(-s)>_0=<\phi(y)\phi(x_s)>_0=:F_{\phi(x)\phi(y)}(s)  \eeq
with $\phi(x),\phi(y)$ replacing $A,B$.
\begin{bem} Due to Lorentz invariance of the vacuum, $\Omega$, we have
\beq <\phi(y)U(s)\phi(x)U(-s)>_0=<\phi(y)U(s)\phi(x)>_0  \eeq
and
\beq <U(s)\phi(x)U(-s)\phi(y)>_0=<\phi(x)U(-s)\phi(y)>_0   \eeq
\end{bem}

We now insert $x_s$ in the expression for $W(x-y)$ getting
\beq \label{1} <\phi(y)\phi(x_s)>_0=\int dk^1/(2\pi 2k^0)\,e^{-ik^0(y^0-x_s^0)}\cdot e^{ik^1(y^1-x_s^1)}  \eeq
Our aim is it to analytically continue this expression by analytically continue $x_s$. This can be done by analytically continue $\cosh s$ and $\sinh s$. With $z:=s+i\mu$ we have
\beq \label{2} x_{s+i\mu}= (x^0\cosh(s+i\mu)+x^1\sinh(s+i\mu),x^1\cosh(s+i\mu)+x^0\sinh(s+i\mu))   \eeq
\begin{bem} At this point one can equally well use Rindler coordinates with $\Lambda(s)(\rho,\eta)=(\rho,\eta+s)$
\end{bem} 

We can insert this expression into the exponent under the integral of the preceding formula, however there is in fact lurking another technical problem in the background. But before studying this problem we go on and choose $\mu=\pi$. By either going back to the original definition for $\sinh x,\cosh x$, i.e.
\beq \sinh x=(e^x-e^{-x})/2\quad \cosh x=(e^x+e^{-x})/2   \eeq
or, using the summation formulas
\beq \sinh(x+y)=\cosh x\cosh y+\cosh x\sinh y\quad \cosh(x+y)=\cosh x\cosh y+\sinh x\sinh y   \eeq
plus
\beq \cosh i\mu=\cos\mu\quad \sinh i\mu=i\sin\mu   \eeq
\beq \sinh(x+i\mu)=-\sinh x\quad \cosh(x+i\mu)=-\cosh x   \eeq

This yields
we get
\beq x_{s+i\pi}=-(x^0\cosh s+x^1\sinh s,x^1\cosh s+x^0\sinh s)=-x_s  \eeq
It follows
\begin{conclusion} We have
\beq F_{\phi(x)\phi(y)}(s+i\pi)=<\phi(y)\phi(-x_s)>_0  \eeq
and, as $y,-x_s$ are spacelike with $y,x_s\in W_R$,
\beq \label{3} <\phi(y)\phi(-x_s)>_0=<\phi(-x_s)\phi(y)>_0   \eeq
\end{conclusion}
\begin{bem}[Warning] At this point we have to spell out a warning. One could have the idea to straightforwardly analytically continue further beyond $s+i\pi$ to e.g. $s+2i\pi$ which would yield
\beq \sinh(s+2i\pi)=\sinh s\quad \cosh(s+2i\pi)=\cosh s  \eeq
 and hence 
\beq F_{\phi(x)\phi(y)}(s+i2\pi)=<\phi(y)\phi(x_s)>_0  \eeq
that is, there would be no sign of thermal i.e. KMS-behavior.
\end{bem}

In order to analyse this problem we have to look more carefully into the analytic continuation of $<\phi(y)\phi(x_s)>_0$. To this end we use the expression of the two-point function as an integral, i.e. formula (\ref{1}) and the following equation (\ref{2}). We make a more explicit calculation of eq. (\ref{2}) and get
\begin{multline} x^0(s+i\mu)=x^0(\cosh s\cos\mu+i\sinh s\sin\mu)+x^1(\sinh s\cos\mu+i\cosh s\sin\mu)\\= \cos\mu\cdot x^0(s)+i\sin\mu\cdot x^1(s)  \end{multline}
\begin{multline} x^1(s+i\mu)=x^0(\sinh s\cos\mu+i\cosh s\sin\mu)+x^1(\cosh s\cos\mu+i\sinh s\sin\mu)\\=\cos\mu\cdot x^1(s)+i\sin\mu\cdot x^0(s)
\end{multline}

When inserting this in formula (\ref{1}) we get an expression under the integral containing a term 
\beq e^{-(k^0x^1(s)-k^1x^0(s))\cdot\sin\mu}  \eeq 
$(x^1(s),x^0(s))$ is a time-like vector in the forward cone for all $s$ if $(x^0,x^1)\in W_R$. Due to the spectrum condition $(k^0,k^1)$ is also a vector in the forward cone, i.e., we have
\beq k^0x^1(s)-k^1x^0(s)>0   \eeq
that is
\begin{conclusion} For $0<\mu<\pi$ the exponent decys strongly and hence the integral is finite. This implies that $<\phi(y)\phi(x_s)>_0$ can be analytically continued into the strip $(s,\mu),<\mu<\pi$ with $F_{\phi(x)\phi(y}(s+i\pi)=<\phi(-x_s)\phi(y)>_0$.On the other hand the integral diverges for $\mu>\pi$ and hence an analytic continuation in this region becomes meaningless in this way.
\end{conclusion}

But we can use formula $(\ref{3})$ and start instead from $<\phi(-x_s)\phi(y)>_0$. By analytically continuing $-x_s=-(\Lambda(s)x)$ to $-x_{s+i\mu}$
we can analytically continue  $<\phi(-x_s)\phi(y)>_0$ into the strip $(s+i\mu),\mu<\pi$ and finally get with $-x_{s+i\pi}=x_s$ the expression $<\phi(x_s)\phi(y)>_0=G_{\phi(x)\phi(y)}(s)$. The corresponding integral exists and is finite by the same reasoning as above. We now piece together these two analytic functions, defining an analytic function in the strip $(s+i\mu),0<\mu<2\pi)$.

We define $F_{\phi(x)\phi(y)}(s+i\mu)$ in the strip $(s+i\mu),0<\mu<\pi$ as above. We define $F_{\phi(x)\phi(y)}(s+i\mu)$ in the strip $(s+i\mu),\pi<\mu<2\pi$ by the analytic continuation of  $<\phi(-x_s)\phi(y)>_0$ as described in the preceding paragraph thus arriving finally at
\beq F_{\phi(x)\phi(y)}(s+i2\pi)=G_{\phi(x)\phi(y)}  \eeq
Our reasoning is complete if we can show that the constructed function is analytic along the line $z=s+i\pi$. Note that up to now analyticity has only been shown in the interior of the two strips. Generally the functions are only continuous at the boundaries of the domains of analyticity. 

To accomplish this we employ the following lemma:
\begin{lemma}[Edge of the Wedge] Let $F_1$ be a function continuous and finite on the closed strip $(z=s+i\mu,0\leq\mu\leq\pi)$ and analytic in the interior. Let $F_2$ be a function having the analogue properties on the closed strip $(z=s+i\mu,\pi\leq\mu\leq 2\pi)$. We assume
\beq F_1(s+i\pi)=F_2(s+i\pi)   \eeq
then $F_1\cup F_2$ extends to an analytic function on the strip $(z=s+i\mu,0<\mu<2\pi)$.
\end{lemma}
Sketch of Proof: Use the Cauchy integral formula both for the interior of the strips belonging to $F_1$ or $F_2$. with part of the boundaries of the Cauchy integrals  being a common interval lying on the line $z=s+i\pi$. These two closed paths can be united to a single closed path by traversing the common interval lying on the line $z=s+i\pi$ in opposite directions. This shows that the extended function $F$ is analytic on the line $z=s+i\pi$.
\begin{conclusion} This proves that the Minkowski vacuum $\Omega$ is a thermal (KMS) state on the observable algebra of the Rindler wedge $W_R$ with $\beta=2\pi$.
\end{conclusion}
\begin{koro} The same result follows of course for the left Rindler wedge $W_L$.
\end{koro}   

We want to conclude this section with a last remark. We emphasize that our version of analytic continuation is essentially unique. We learned that the direct analytic continuation, starting from $<\phi(y)\phi(-x_s)>_0$ beyond $(\mu\leq\pi)$ would lead to a divergent integral. But let us assume this method would work. Then we had apparently two different analytic continuations, starting either from $<\phi(y)\phi(-x_s)>_0$ or $<\phi(-x_s)\phi(y)>_0$. But both continuations coincide on the boundary (as we have seen). Hence the difference of these continuations is analytic in the upper strip and vanishes on the boundary $(z=s+i\pi)$. With the help of the Schwartz- reflection priciple we then would get an analytic function which vanishes on an interior line. Hence the difference of the two functions vanishes in its domain of analyticity and we see that the two functions have to coincide.
\section{The Universal Structure of Thermal States as a System built from Quasiparticles and Holes and its Relation to Rindler Space}
In this section we want to show that all the different observations being made in connection with the Unruh effect have a, in or view, common physical origin, that is, the emergence of a new kind of \tit{creation/annihilation operators} in thermal systems of many DoF. The preceding analysis has shown that an example is given by the right or left Rindler wedge, $W_R,W_L$.

A thermal system is full of elementary excitations. It was an ingeneous insight of Landau to replace the original system with its ordinary microscopic DoF by a better adapted choice of DoF, that is, the so-called \tit{elementary or collective excitations}, which are adapted to the Hamiltonian of the system insofar as they are assumed to interact weakly and approximately diagonalize the complex Hamiltonian. This was discussed in more detail in \cite{NRT} and in the follow-up paper \cite{Requ2}. A related point of view has been adopted in the so-called thermofield theory of Umezawa et al. (see e.g. \cite{Umezawa}). We recently discussed this topic in quite some detail in \cite{Requ1}. It should however be emphasized that in most representations this form of doubling or extension, we will develop in the following, is delineated as a purely formal calculational technique.

It is remarkable that this formal structure was already observed much earlier in \cite{Araki}, again without giving a physical interpretation. A certain exception is however \cite{Israel}. We observed and discussed some aspects of this phenomenon in our doctoral thesis (\cite{Goettingen} or \cite{Requ3}). More specifically, we analyzed the intriguing symmetry properties of the Fourier spectrum of correlation functions in a thermal KMS-state. Choosing, for convenience, $A,B$ selfadjount, the KMS condition   
\beq <A(t)B>=<BA(t+i\beta)>   \eeq
(where we are a little bit sloppy in the precise definition of $A(t+i\beta)$) can be rewritten as 
\beq <A(t)B>-<BA(t)>=<[A(t),B]>=<B(A(t+i\beta)-A(t))>   \eeq
With $J(\omega)$ the Fourier transform of $(<BA(t)>-<A><B>)$ and $C(\omega)$ the Fourier transform of $<[A(t),B]>$, i.e.,
\beq <BA(t)>-<A><B>=(2\pi)^{-1/2}\int e^{-it\omega}J(\omega)\,d\omega  \eeq
we get the relation
\beq C(\omega)=(e^{\beta\omega}-1)J(\omega)   \eeq
\begin{bem} As it sometimes happens that the commutator $[A(t),B]$ is relatively simple (e.g. a c-number) we can get with the help of this formula a quite explicit expression for $<BA(t)>$. 
\end{bem}

With
\beq \overline{<\Omega|A(t)B\Omega>}=<A(t)B\Omega|\Omega>=<\Omega|BA(t)\Omega>  \eeq
that is
\beq <\Omega|A(t)B\Omega>= \overline{<\Omega|BA(t)\Omega>}  \eeq
we get for the respective Fourier transforms:
\beq F.Tr.(<A(t)B>-<A><B>)=\overline{J(-\omega)}  \eeq
This yields for the commutator:
\beq \overline{J(-\omega)}-J(\omega)=(e^{\beta\omega}-1)J(\omega)  \eeq
that is
\beq  \overline{J(-\omega)}=e^{\beta\omega}J(\omega)   \eeq
and hence
\begin{ob} From the KMS-condition it follows
\beq Re\,J(-\omega)=e^{\beta\omega}\,Re\,J(\omega)\quad Im\,J(-\omega)=-e^{\beta\omega}\,Im\,J(\omega)   \eeq
\end{ob}
\begin{bem} Note that in the above cited papers we used a different sign convention in the Fourier transform.
\end{bem}
\begin{conclusion} Due to the above inherent symmetry of the spectrum of the KMS Hamiltonian the existence of a (quasi-) particle branch for positive $\omega$ implies the existence of a (quasi-) hole excitation branch for negative $\omega$. This picture becomes even more transparent for translation invariant systems, the situation we discussed in the above mentioned literature, where we have Fourier transforms in both $\omega$ and $\mbf{k}$ and hence true excitation branches (see also \cite{NRT}, where this picture played a central role).
\end{conclusion}
 
It is clear that a (strongly) interacting system of many DoF (like a quantum fluid) is not really completely equivalent to a system of non-interacting collective excitations. But this picture is supposed to hold for the low-lying elementary excitations and sufficiently low temperatures. In our case of the Unruh effect the system is assumed to be a free KG-system anyway. That is, the above picture should hold in this case (and it holds also approximately for interacting systems (as we have shown in \cite{NRT} and \cite{Requ2}).  That is, we begin our discussion by developing the picture for a general free thermal quantum system. In a next step we apply our results to the Unruh scenario.

Another delicate point we want to address in this context is the problematical relation of the Minkowskian and Rindlerian point of view when comparing what happens in $W_R$ or $W_L$ as both frameworks employ different and (possibly inequivalent) Hilbert spaces. This crucial point is frequently glossed over in the existing discussions. In the Rindler case we have the Rindler Fock space and a temperature state (assumed to represent the Minkowski vacuum) which is usually modelled as a density matrix over the Rindler Fock space. This turns out to be grossly inadequate and we will replace it by a true infinite thermal (KMS) state which we then compare with the original Minkowski vacuum in $W_R$ or $W_L$.
\subsection{Creation/Annihilation Operators for Collective Excitations}
We adapt our notation to the present situation of the Unruh-Rindler scenario by not using indices which (usually) denote momentum eigenvectors parametrized by the index $\mbf{k}$. In our case translation invariance is absent and we have instead an eigenfunction expansion with respect to an energy label denoted by $\omega$, the Fourier variable belonging to the Rindler time $\eta$, given by the infinitesimal generator of Lorentz boosts (for more details see the following subsection). For the sake of brevity our starting point is a physical one. That this working philosophy is correct will then be seen below and has also been laid out in \cite{NRT},\cite{Requ1} and \cite{Requ2}.

We assume a thermal quantum system be given (for convenience we restrict ourselves for the time being to a free Bose system as in the Unruh case). The ordinary particle annihilation/creation operators are denoted by $a(\omega),a^+(\omega)$. Our physical input is derived from the following crucial observation. The ordinary particle annihilation/creation operators in a thermal state are conjectured to consist of two pieces which are difficult to observe individually but occur in the following temperature dependent superposition.
\begin{ob} We conjecture the following universal splitting to hold in a thermal state:
\beq a(\omega)=(1+f_{\beta}(\omega))^{1/2}a(\omega,\beta)+f_{\beta}^{1/2}\tilde{a}^+(\omega,\beta)   \eeq
\beq  a^+(\omega)=(1+f_{\beta}(\omega))^{1/2}a^+(\omega,\beta)+f_{\beta}^{1/2}\tilde{a}(\omega,\beta)   \eeq
The concrete functional shape of the positive function $f_{\beta}(\omega)$ will be calculated below. The $a(\omega,\beta),a^+(\omega,\beta),\tilde{a}(\omega,\beta),\tilde{a}^+(\omega,\beta)$ are quasi-particle annihilation/creation operators, quasi-hole annihilation/creation operators, respectively.
Their crucial property is that $a(\omega,\beta),\tilde{a}(\omega,\beta)$ annihilate the  thermal state $\Omega_{\beta}$ expressed as a Hilbert vector in some thermal Hilbert space. This is in marked contrast to the 'real' operators $a(\omega)$!
\end{ob}

The physical motivation underlying this representation is the following. In contrast to a ground state a thermal state supports many excitations which can be regarded either as quasi- particle excitations or as excitations of holes in the already existing distribution of real particles. For example, a 'real' particle annihilation operator can be regarded as a superposition of a quasi-particle annihilation operator and the creation of a hole in the existing sea of particles. It is remarkable that these new and somewhat hidden excitation modes are temperature dependent in contrast to the original particle annihilation/creation operators. 
\begin{bem} In \cite{Requ1} we discussed this phenomenon in the context of the old Dirac picture. Furthermore, at the moment we do not strictly distinguish between elementary excitations, collective excitations or quasi particles which are considered in some of the existing literature as different modes of excitation.
\end{bem}
 
We assume that these thermal (quasi-) particle/hole operators fullfil canonical commutation relations, i.e.
\beq [a(\omega,\beta),a^+(\omega',\beta)]=\delta(\omega-\omega')  \eeq
\beq  [\tilde{a}(\omega,\beta),\tilde{a}^+(\omega',\beta)]=\delta(\omega-\omega')  \eeq
\beq [a(\omega,\beta),\tilde{a}^+(\omega',\beta)]= 0 \eeq
with the remaining combinations vanishing identically.
This yields
\beq [a(\omega),a^+(\omega')]=(1+f_{\beta})\delta(\omega-\omega')-f_{\beta}\delta(\omega-\omega')=\delta(\omega-\omega')  \eeq
that is, corresponding canonical commutation relations follow for the real annihilation/creations operqators.

The concrete functional form of $f_{\beta}(\omega)$ can be inferred from the condition that in a thermal state we want to have 
\beq <a^+(\omega)a(\omega')>_{\beta}=(\Omega_{\beta}|a^+(\omega)a(\omega')\Omega_{\beta})=(e^{\beta\omega}-1)^{-1}\cdot \delta(\omega-\omega')  \eeq
I.e., the well-known occupation number in the Bose case. Inserting the above superpositions in this formula we get with 
\beq  a(\omega,\beta)\Omega_{\beta}=0= \tilde{a}(\omega,\beta)\Omega_{\beta}  \eeq
\begin{multline} (\Omega_{\beta}|a^+(\omega)a(\omega')\Omega_{\beta})=(\Omega_{\beta}|f_{\beta}^{1/2}(\omega)\tilde{a}(\omega,\beta)\cdot f_{\beta}^{1/2}(\omega')\tilde{a}^+(\omega,\beta)\Omega_{\beta})\\ =f_{\beta}^{1/2}(\omega)f_{\beta}^{1/2}(\omega')(\Omega_{\beta}|[\tilde{a}(\omega,\beta),\tilde{a}^+(\omega',\beta]\Omega_{\beta})=f_{\beta}(\omega)\delta(\omega-\omega')
\end{multline}
That is, we have
\begin{lemma} It holds
\beq f_{\beta}(\omega)=(e^{\beta\omega}-1)^{-1}\quad\text{and}\quad 1+ f_{\beta}(\omega)=e^{\beta\omega}/(e^{\beta\omega}-1)   \eeq
\end{lemma}

It is remarkable that from these fundamental thermal annihilation/creation operators we can construct another real representation which commutes with the above real representation. Defining
\beq\tilde{a}(\omega)=(1+f_{\beta}(\omega)^{1/2}\tilde{a}(\omega,\beta)+f_{\beta}^{1/2}a^+(\omega,\beta)   \eeq
we have for example
\begin{multline} [\tilde{a}(\omega),a^+(\omega')]=(1+f_{\beta})^{1/2}f_{\beta}(\omega')^{1/2}[\tilde{a}(\omega,\beta),\tilde{a}(\omega',\beta)]\\+f_{\beta}^{1/2}(\omega)(1+f_{\beta}(\omega')^{1/2}[a^+(\omega,\beta),a^+(\omega',\beta)]=0  \end{multline}
and correspondingly for the other combinations.
\begin{ob} From the thermal creation/annihilation operators we can construct a tilde representation $(\tilde{a}(\omega),\tilde{a}^+(\omega))$ which commutes with the $(a(\omega),a^+(\omega))$ representation with
\beq\tilde{a}(\omega)=(1+f_{\beta}(\omega)^{1/2}\tilde{a}(\omega,\beta)+f_{\beta}^{1/2}a^+(\omega,\beta)   \eeq
\end{ob}
We see that in the tilde representation the notion of thermal particles and holes are exchanged.

We can invert the above expressions. From
\beq \begin{pmatrix} a(\omega) \\ \tilde{a}^+(\omega) \end{pmatrix}=\begin{pmatrix} (1+f)^{1/2} & f^{1/2} \\ f^{1/2} & (1+f)^{1/2} \end{pmatrix}\,\begin{pmatrix} a(\omega,\beta) \\\tilde{a}^+(\omega,\beta) \end{pmatrix}  \eeq
we get
\beq \begin{pmatrix} a(\omega,\beta) \\\tilde{a}^+(\omega,\beta) \end{pmatrix}=\begin{pmatrix} (1+f)^{1/2} & -f^{1/2} \\ -f^{1/2} & (1+f)^{1/2}\end{pmatrix}\, \begin{pmatrix} a(\omega) \\ \tilde{a}^+(\omega) \end{pmatrix}  \eeq
that is
\begin{ob} The thermal particle/hole operators are expressed as
\beq a(\omega,\beta)=(1+f_{\beta})^{1/2}a(\omega)-f_{\beta}^{1/2}\tilde{a}^+(\omega)  \eeq
\beq \tilde{a}(\omega,\beta)=-f_{\beta}(\omega)^{1/2}a^+(\omega)+(1+f_{\beta})^{1/2}\tilde{a}(\omega)  \eeq
and correspondingly for the operators $a^+(\omega,\beta),\tilde{a}^+(\omega,\beta)$.  
\end{ob}

The observation that there does exist kind of a symmetry or duality between the v.Neumann algebra of observables $\mcal{A}$ and its commutant $\tilde{\mcal{A}}$ in a thermal representation in some Hilbert space is a well-known abstract structural phenomenon (see e.g. \cite{Bratteli} and \cite{Bratteli2}). We would however like to add some physical remarks which are frequently missing in the more general analysis and which put some flesh on the abstract structure. Furthermore it exhibits in our view an existing physical universal deep structure lying beneath the abstract formalism.

In our case we deal primarily with infinitely extended structures. This implies that simple explanations which rely for example on tools like density matrices etc. are not really helpful. There do exist several possibilities to motivate the existence of such an apriori structure as described above. The one which leads rather immediately to the well-known tensorial double structure we observe in the black hole or Unruh scenario (and which is e.g. formalized in the papers by Kay et al., see for example \cite{Kay1} and \cite{Kay2}), is based on the method to construct a thermal state by tracing over another Hilbert space, viewed as a tensor factor. In that case the corresponding thermal Hilbert space is the tensor product of two tensor factors on which dual pictures of the original observable algebra are realized.

We want in the following to follow a slightly different line of ideas as they do lead us more directly to the fundamental and slighly hidden structure underlying this field and which we described above. The main problem in constructing a thermal vectorn state, representing a thermal eqilibrium state, is to cope with the (infinite) fluctuation energies which occur in the representation if we are going to perform the termodynamic limit. That is, we have both to renormalize the average energy of the equilibrium state which tries to evade to infinity and the energy fluctuations which also will diverge. More specifically, we assume the existence of a \tit{typical} thermal vector state which support a great number of ordinary (real) excitations and which are distributed in such a way that all 'holes' below the so-called 'Fermi surface' are occupied and no quasi-particle excitations above the Fermi surface are excited.

Furthermore the energy of the Fermi surface is reset to zero. More precisely, the quasi-particle and hole annihilation operators annihilate this 'thermal vacuum'.
\begin{bem} In a sense this picture, we are envoking, is reminiscent of the old Dirac picture which, however, appears to be more justified in this thermal context. In this context the notion Fermi surface also is making some sense.
\end{bem} 
\begin{ob} In this picture it is now possible to reinterpret the meaning of the respective annihilation and creation operators by introducing a certain dualization symmetry. I.e., by exchanging the meaning of quasi-particles and holes we get another algebra of observables called $\tilde{\mcal{A}}$ which commutes with the ordinary algebra $\mcal{A}$.
\end{ob} 

This dual structure becomes possible by taking the concept of quasi-particles and holes as the really fundamental strucuture, while the ordinary field operators and observables in the thermal state become certain superpositions of these more primordial objects. On the other hand, as quasi-particle and hole operators do commute, they generate automatically a certain tensor product structure of two Fock spaces of, on the one hand, quasi-particles lying above the Fermi surface and, on the other hand, holes lying below the Fermi surface.
\begin{ob} This physically motivated tensor product structure leads quasi automatically to the mathematically motivated structure found by Araki et al. (\cite{Araki}).
\end{ob}
\subsection{Constructing the v.Neumann Observable Algebras of the Left/Right Wedge in the Thermal Rindler Hilbert Space}
We have learned in section 2 that the observable algebras of $W_R,W_L$ in Minkowski space, (that is, formulated with the help of ordinary Minkowski space-time coordinates) are not identical to the coresponding algebras if expressed by means of Rindler coordinates and represented in Rindler Fock space. More specifically, the respective v.Neumann algebras (i.e. the weak or ultra-weak closures) behave differently concerning their limit behavior when approaching the boundaries (i.e. $t=\pm x$). 

This is the reason why the Minkowski vacuum cannot be a density matrix over the Rindler Fock space as we have proved in section 2. As a consequence many of the calculations in most of the literature about the Unruh effect, if based on this unjustified identification and its ramifications, do have only a heuristic meaning. In this subsection we want to show how this ambiguity can be remedied.

Recapitulating the abstract results we mentioned in section 2, we see that we would get a weakly continuous $\ast-$ isomorphism between the two observable algebras, on the one hand on the two wedges in Minkowski space, on the other hand on Rindler space, if the limit construction works in both cases. This would entail that the class of pure states and density matrices (called the folium) is the same in both cases.    

We will show in the following that in contrast to the Rindler Fock space the situation is much better when dealing with the thermal  (KMS) state over the Rindler Fock space as described in the preceding section. Again we have to scrutinize the behavior of limit elements when approaching the boundaries of $W_R,W_L$ in Rindler space. Somewhat surprisingly, it turns out that, in contrast to ordinary Rindler Fock space, the existence of limit elements can be proved as a consequence of a mathematical subtlety.

Before entering into the proof we want to give a more detailed account of the behavior of the support of observables when we approach the boundary of the wedge. As in section 2 we assume the support of the sequence of observables or fields to be concentrated between the planes or lines 
\beq  t+x=a>0\quad , \quad t+x=b>a    \eeq
while the support shrinks in the transverse direction $t-x$ as $-n^{-1}$, that is, in the coordinate $t-x$ we again use a sequence of functions
\beq h_n(t-x)\quad ,\quad \int h_n(s)ds=1 \quad ,\quad \text{for eample}\quad  h_n(s)=n\cdot h(ns)  \eeq 
We illustrate the behavior with the help of a sequence of functions
\beq F(t,x)=f(t+x)\cdot h_n(t-x)   \eeq
with $supp\, h_n$ shrinking to zero for $n\to\infty$.

We want to analyze the support with respect to the corresponding Rindler coordinate $\eta$ which goes to $\infty$ for $t-x\to 0$ because this detailed behavior becomes relevant for our following analysis. We study the behavior for the endpoints of the interval, i.e.
\beq t+x=a \quad ,\quad t+x=b   \eeq
We have
\beq t+x=\rho(\sinh\,\eta+\cosh\,\eta)=\rho e^{\eta}  \eeq
\beq t-x=-n^{-1}=\rho(\sinh\,\eta-\cosh\,\eta)=-\rho e^{-\eta}  \eeq
That is
\beq a=\rho_a e^{\eta_a}\quad ,\quad b=\rho_b e^{\eta_b}   \eeq
and
\beq \rho_a=n_a^{-1}e^{\eta_a}\quad ,\quad \rho_b=n_b^{-1}e^{\eta_b}  \eeq
which yields
 \beq  a=n^{-1}e^{2\eta_a} \quad ,\quad b=n^{-1}e^{2\eta_b}\quad\text{or}\quad na= e^{2\eta_a}\quad ,\quad nb= e^{2\eta_b}  \eeq

We want to calculate $(\eta_a-\eta_b)$ for $n\to\infty$. We have
\beq\ln\,n+\ln\,a=2\eta_a\quad ,\quad \ln\,n+\ln\,b=2\eta_b  \eeq
and hence
\beq \eta_a-\eta_b=(\ln\,a-\ln\,b)/2   \eeq
\begin{lemma} While $\eta_{a,b}$ go to $\infty$ with $n\to\infty$, their difference remains bounded. This entails that the support of testfunctions concentrate at $\rho=0$ and remains bounded with respect to $\eta$ while being shifted to $\eta\to\infty$.
\end{lemma}

In the following analysis only the assymptotic behavior of the $\eta-$ dependence for $\eta\to\infty$ is relevant. The dual variable of $\eta$ is $\omega$ which we introduced in the preceding section. We want to show that a combination of $\omega$-dependent prefactors of the respective quasi-particle/hole creation/annihilation operators in Rindler space conspire in making the $\eta\to\infty$ limit smooth enough so that a limit operator can actually be defined. As v.Neumann algebras are both strongly and weakly closed we will smear the Rindler field operators with n-dependent testfunctions and assume that they are applied to fixed Rindler Hilbert space vectors and show that these limits do exist for $n\to\infty$.

In a first step we realize that in the Fourier expansion of fields in Rindler space there occurs a prefactor $\omega^{-1/2}$. Furthermore the thermal creation/annihilation operators carry prefactors of the type $(e^{\beta\omega}-1)^{-1/2}$. We now choose a testfunction $h_n(\eta)$ having its support in the interval $(\eta_a(n),\eta_b(n))$ as discussed above. Its Fourier transform 
\beq  (2\pi)^{-1/2}\int e^{i\omega\eta}\cdot h_n(\eta)\,d\eta   \eeq
behaves in the following way for $n\to\infty$

For convenience we assume that $h_n$ is centered around some $\eta_n\in (\eta_a(n),\eta_b(n))$ in the form 
\beq  h_n(\eta)=h(\eta-\eta_n)   \eeq
This yields
\begin{multline}  (2\pi)^{-1/2}\int e^{i\omega\eta}\cdot h_n(\eta)\,d\eta= (2\pi)^{-1/2}\int e^{i\omega(\eta+\eta')}\cdot h(\eta')\,d\eta' = \\
e^{i\omega\eta_n}\cdot (2\pi)^{-1/2}\int e^{i\omega\eta'}\cdot h_(\eta')\,d\eta'=e^{i\omega\eta_n}\cdot \tilde{h}(\omega)  \end{multline}

We now consider the individual terms which occur in the ordinary Rindler field operators, i.e., the quasi-particle/hole creation/annihilation operators smeared with the $e^{i\omega\eta_n}$ plus the extra factors $\omega^{-1/2}\cdot (e^{\beta\omega}-1)^{-1/2}$.
For $\eta$ or $\eta_n$ large or going to infinity only an infinitesimal neighborhood of $\omega=0$ is relevant in the integral
\beq \int \omega^{-1/2}\cdot (e^{\beta\omega}-1)^{-1/2}\cdot e^{i\omega\eta_n}\tilde{h}(\omega)\cdot A(\omega)\,d\omega  \eeq
with $A(\omega)$ representing a quasi-particle/hole creation/annihilation operator. Asymptotically for $n\to\infty$ we thus have
\beq \int \omega^{-1}\cdot e^{i\omega\eta_n}\tilde{h}(\omega)\cdot A(\omega)\,d\omega  \eeq

We see that 
\beq -i\,d/d\eta_n\int  \omega^{-1}\cdot e^{i\omega\eta_n}\tilde{h}(\omega)\cdot A(\omega)\,d\omega =\int   e^{i\omega\eta_n}\tilde{h}(\omega)\cdot A(\omega)\,d\omega \eeq
Noting that we assumed this expression to be applied to some vector and applying the Riemann-Lesbegue lemma, we see that for $\eta_n\to\infty$ this expression goes to zero. In other words, the expression
\beq \int \omega^{-1}\cdot e^{i\omega\eta_n}\tilde{h}(\omega)\cdot A(\omega)\,d\omega \cdot \psi \eeq
with $\psi$ some arbitrary Hilbert space vector, converges asymptotically for $\eta_n\to\infty$ towards a constant vector in the thermal Rindler Hilbert space.
\begin{conclusion} We conclude that the strong limits of observables as described above do exist if they approach the boundary $(\rho\to\ 0,\eta\to\infty)$. This entails that the strong closure of the original unclosed algebras of observables, having their support in the open interior of $W_R$ or $W_L$ do exist and we see that we get a one-one correspondence of the respective v.Neumann algebras over the left/right wedge in the thermal Rindler Hilbert space $\mcal{H}^{th}_R$ and the corresponding v.Neumann algebras over the left/right wedge in Minkowski Hilbert space $\mcal{H}_M$.
\end{conclusion}
\begin{bem} We note that these v.Neumann algebras are closed in most of the other topologies as e.g. weak, $\sigma$-weak etc.
\end{bem} 
\section{The Passage from Rindler to Minkowski Space}
We learned from the construction in the preceding section that the v.Neumann observable algebras of the respective wedges, $W_L,W_R$ in Minkowski space $\mcal{H}_M$ are in one-one correspondence to the correspondng algebras  in the thermal Rindler Hilbert space $\mcal{H}^{th}_R$ . We denote these v.Neumann algebras by
\beq \mcal{A}(W_L),\mcal{A}(W_R)\quad\text{in Minkowski space}\quad \mcal{B}(W_L),\mcal{B}(W_R)\quad\text{in Rindler space}  \eeq

As described in section 2 it follows that the algebraic isomorphism is $\sigma$-weakly continuous in both directions and that the sets of normal states are identical. This then holds also for their respective unifications, i.e. the v.Neumann algebras generated by the union of left/right algebras, denoted by
\beq  \mcal{A}(W_L)\vee\mcal{A}(W_R)\quad\text{and}\quad\mcal{B}(W_L)\vee\mcal{B}(W_R)  \eeq

Usually we can assume that the v.Neumann algebras
\beq \mcal{A}(W_L),\mcal{A}(W_R)\quad\mcal{B}(W_L),\mcal{B}(W_R)  \eeq
are factors, i.e., they have trivial centers $\{\lambda\cdot\mbf{1}\}$. Furthermore, the algebras of the left wedge are the commutants of the algebras of the right wedge and vice versa, that is
\beq \mcal{A}(W_L)=\mcal{A}(W_R)'     \quad\mcal{B}(W_L)=\mcal{B}(W_R)'   \eeq
From this follows immediately:
\begin{conclusion} $ \mcal{A}(W_L)\vee\mcal{A}(W_R)$ and $ \mcal{B}(W_L)\vee\mcal{B}(W_R)$ are irreducible on their respective Hilbert spaces, that is, they comprise all bounded operators.\\
Furthermore, all these algebras have a cyclic and separating vector, that is, the Minkowski vacuum in the Minkowski case, the thermal Rindler vacuum vector in the Rindler case.\\
From both properties follows that the weakly continuous isomorphism $\Phi$ defined above is actually unitarily implementable and $U$ maps the thermal vacuum vector $\Omega^{th}_R$ onto the Minkowski vaccum vector $\Omega_M$.
\end{conclusion}
Proof: The famous double commutant theorem of v.Neumann implies that 
\beq (\mcal{A}(W_L)\vee\mcal{A}(W_R))= (\mcal{A}(W_L)\vee\mcal{A}(W_R)''=\{\lambda\cdot\mbf{1}\}'  \eeq
\beq   (\mcal{B}(W_L)\vee\mcal{B}(W_R)= (\mcal{B}(W_L)\vee\mcal{B}(W_R)''=\{\lambda\cdot\mbf{1}\}'  \eeq
with the rhs being the algebras of all bounded operators. The second property is well known.
From this follows the unitary implementability, see e.g. \cite{Bratteli}.\vspace{0.3cm}
\begin{bem} We see a relation between this result, i.e., that we get the full algebra of boundd operators in $\mcal{H}_M$, and the possibility of finding complete systems of mode expansions in Minkowski Hilbert space of Rindler modes via analytic continuation, as described e.g. in \cite{U}.
\end{bem} 

We can now use these findings to answer our questions we have raised in the introduction. With $U\circ U^{-1}$ all the expressions occurring in $\mcal{H}^{th}_R$ can be transferred to $\mcal{H}_M$. For the KG-field itself we have
\beq U\phi_R(\rho,\eta)U^{-1}=\phi_M(x(\rho,\eta))  \eeq
As both fields fulfill the KG-equation we can employ the KG-scalar product with respect to the various mode expansions (Minkowski or Rindler modes) to generate the respective operator mode expansions in Minkowski Hilbert space $\mcal{H}_M$. For the ordinary Rindler mode operators we get
\beq Ua^{(+)}_RU^{-1}=:a^{(+)}_{R/M}\quad ,\quad  U\tilde{a}^{(+)}_RU^{-1}=:\tilde{a}^{(+)}_{R/M}  \eeq
with the rhs being the corresponding images of the Rindler operators in Minkowski Hilbert space (belonging to $W_R$ or $W_L$) The concrete transformation formulas (Bogoliubov expansion) we find e.g. in \cite{F} or the other cited papers. They remain correct as they are of a purely algebraic character. One should however note that the underlying physics is nevertheless questionable as they usually start from the Rindler Fock space which is incorrect as we showed above.

Interesting is the role of the images of the fundamental quasi-particle/hole creation/annihilation of thermal Rindler Hilbert space in Minkowski Hilbert space. As $a(\beta,\omega),\tilde{a}(\beta,\omega)$ annihilate $\Omega^{th}_R$, we have
\beq Ua(\beta,\omega)U^{-1}\Omega_M=  U\tilde{a}(\beta,\omega)U^{-1}\Omega_M=0 \label{99}   \eeq
We hence get
\begin{conclusion} $ Ua(\beta,\omega)U^{-1}\,,\, U\tilde{a}(\beta,\omega)U^{-1}$  are suitable superpositions of \\Minkowski-annihilation operators. On the other hand,  $a(\beta,\omega),\tilde{a}(\beta,\omega)$ are superpositions of the ordinary creation/annihilation operators in $\mcal{H}^{th}_R$ (see section 4.1), which are mapped onto $a^{(+)}_{R/M}\,,\,\tilde{a}^{(+)}_{R/M}$. These latter operators can be expressed by superpositions of Minkowski creation/annihilation operators as we described above. Therefore this yields, by the same token, explicit expressions in Minkowski Hilbert space for $ Ua(\beta,\omega)U^{-1}\,,\, U\tilde{a}(\beta,\omega)U^{-1}$.
\end{conclusion}

It is interesting that we find analogous expressions in the literature, whereas the canonical framework, as we described above, is rather different. Formula (\ref{99}) says that the operators $Ua(\beta,\omega)U^{-1}, U\tilde{a}(\beta,\omega)U^{-1}$ annihilate the Minkowski vacuum. Originally these are quantum modes stemming from the thermal Rindler Hilbert space framework. In \cite{U} in the formula (2.19a) or in \cite{C} formulas (2.66,2.67) similar properties are expressed while in these approaches the quasi-particle/hole creation/annihilation operators of the thermal Rindler Hilbert space do not openly exist. These papers are rather based on the relation of Rindler Fock space to Minkowski Hilbert space, a relation we think, is debatable, to say the least.
\section{Conclusion}
We have shown in this paper that the appropriate dual quantum field theory to the Minkowski quantum field theory in $W_R$ or $W_L$ in the Unruh scenario is not the Rindler Fock space theory but the thermal quantum field theory defined on the thermal Rindler Hilbert space $\mcal{H}^{th}_R$. 
In contrast to Rindler Fock space, this thermal field theory is unitarily equivalent to the theory in Minkowski space. We showed in particular that under this unitary map the thermal Rindler vacuum is mapped onto the Minkowski vacuum. This thermal Rindler vacuum, which is unitarily related to the Minkowski vacuum, replaces the formal (but incorrect) representation of the Minkowski vacuum as a superposition of certain Rindler modes based on Rindler Fock space in the standard literature.

In the course of the construction of this duality we proved the existence of quasi-particle/hole creation/annihilation operators in the thermal Rindler Hilbert space which turned out to be the fundamental objects in this framework. Under the unitary map $U$ they are mapped into operators living in Minkowski Hilbert space which annihilate the Minkowski vacuum and generate a new mode expansion in Minkowski Hilbert space. The details of the construction shows that the Rindler particles are not really a new kind of particles but are rather a new class of mode representation built from the original Minkowski particles.

\end{document}